# Cloud ArcGIS Online as an innovative tool for developing geoinformation competence with future geography teachers

Ihor V. Kholoshyn[1][0000-0002-9571-1758], Olga V. Bondarenko[1][0000-0003-2356-2674],
Olena V. Hanchuk[1][0000-0002-3866-1133] and Ekaterina O. Shmeltser[2]

[1] Kryvyi Rih State Pedagogical University, 54, Gagarina Ave., Kryvyi Rih, 50086, Ukraine
{holoshyn, bondarenko.olga}@kdpu.edu.ua, elena.ganchuk@gmail.com
[2] Kryvyi Rih Metallurgical Institute of the National Metallurgical Academy of Ukraine,
5, Stepana Tilhy Str., Kryvyi Rih, 50006, Ukraine

**Abstract.** The article dwells upon the scientifically relevant problem of using cloud-based GIS-technologies when training future geography teachers (based on ArcGIS Online application). The authors outline the basic principles for implementing ArcGIS Online in the educational process (interdisciplinary integration, the sequence of individualization in training, communicability, distance education and regional studies), and provide an example of an interactive map created with the help of the specified cloud GIS, since this kind of map is the most popular a form of research by geography students. In the article it is noted that integration of ArcGIS Online into the educational process allows the teacher to follow a clear pedagogical strategy, taking into account possible variants of its use (demonstration, direct mastering of GIS in a computer class and independent work in an individual mode). Considering cloud GIS as a new stage in the development of geoinformational education, the authors emphasize their key benefits (round-the-clock access, work with GIS package in the cloud, the ability to use other maps as well as the creation of their own maps and web-applications) and disadvantages (monetization of services, underestimation of the GIS role in the curriculum of the higher school, the lack of Ukrainian content, etc.).

## 1 Introduction

### 1.1 Scientific relevance of the research

Nowadays, for most educators, it is clear that GIS-based training should take its decent place in the educational process as well as methods of learning about the world (in all time and space scales); as a method of solving actual natural or social problems and as a way of presenting spatial disproportions in the processes and phenomena development. However, as it is stated in the report of the National Research Council of Great Britain (NRC), for the successful implementation of GIS into training, five interrelated conditions are to be ensured, including: financial support, technical support, methodological assistance, GIS introduction into the curriculum and support of community [17]. Note that the provision of these conditions during the study of geography has a number of difficulties that must be overcome by the teacher.

In the context of the above mentioned, it is worth noting the large-scale work being done by ESRI on interactive teaching of educators in the field of GIS-education of various levels and profiling. In particular, a great deal of attention is paid to the introduction of educational materials based on the ArcGIS Online cloud system. In some countries in Europe and the world, cloud-based GIS is already being actively used in the educational process, starting with secondary school education.

### 1.2  Recent research and publications analysis

The analysis of scientific literature on the problem we are investigating convincingly testifies that many works of domestic and foreign authors are devoted to the use of cloud technologies in geoinformational education. Thus, studies by Olga V. Bondarenko [5], Jack Dangermond [6], Iryna M. Khudiakova [8; 9], Svitlana V. Mantulenko [3], Oksana M. Markova [10], Oleksandr V. Merzlykin [12], Vladimir S. Morkun [15], Pavlo P. Nechypurenko [16], Serhiy O. Semerikov [13], Kateryna I. Slovak [14], Andrii M. Striuk [11], Tetiana V. Zaitseva [19], Vladimir I. Zaselskiy [4] etc., have proved that today, in order to improve the learning process, powerful technologies such as cloud computing are needed that, by supporting traditional forms of education, they are a new stage in the development of education and an economically viable, effective and flexible way to meet the needs of those who are being trained to acquire new knowledge.

In the writings of Dmytro S. Zanko [18], Oleksandr V. Barladin [2], Witold Lenart [17] etc. practical tips and examples on using cloud GIS in higher education institutions while teaching a range of professional-cycle subjects (cartographers, informants, etc.) are provided. However, until now, there is no clarity as to how to integrate cloud technologies into the daily teaching of future geography educators.

### 1.3  Article objective

The aim of the proposed study is the theoretical substantiation of the principles and methods of using cloud GIS, in particular ArcGIS Online, for the formation of geoinformation competence in future geography teachers' training.

## 2  Presenting main material

Geographic information systems in geographic education began their way from the late 1970s to the university level and interest in them grew in an avalanche. So, the number of geographic curriculum programs that the American and Canadian universities launched in 1984 equaled about 10. By the end of 1990, their number exceeded 2000, and the use of the field has expanded at the expense of history, computer science, biology, mathematics and other sciences [7].

Pioneers in the use of GIS during the higher learning of geography have become the Harvard University Computer Graphics Laboratory (the USA) and the Department of Experimental Cartography at the Royal College of Arts (the UK). In these institutions, new computer algorithms and programs designed to handle spatial information were developed through attempts and experiments. Thanks to this, active students and teachers have gained the first experience of integration of GIS technologies into the educational process, which has become actively distributed among other world institutions of higher education.

In particular, in 1995 a joint project of hundreds of American universities "Ecological and Spatial Technologies" (EAST) was launched, which used the strategy and technology of problem learning to stimulate the intellectual development of students based on GIS. Since then, GIS technology has begun to develop so smoothly that educational opportunities simply did not have time for them. GIS software products began to generate profits and were quickly bought off.

In the mid-1990s, it became clear that there were not enough specialists able to carry basic geoinformational knowledge to future teachers. This was due to the development of the "Project for the training of GIS NCGIA" by the National Science Foundation of the United States. This project was based on the premise that the educational materials being developed would be widely distributed among educators teaching GIS. The core of the course, about 1000 pages, has been acquired by many educational institutions in the world (over 5 years more than 70 countries had about 1300 copies of the study course). The project is translated into many languages of the world. Until the 1990s, according to the international survey, more than 450 universities in the USA, Europe and Australia were registered providing the opportunity for GIS education [1]. As a result, partnership agreement between schools and higher education institutions contributed greatly to solving the problems of primary GIS education.

The active and comprehensive introduction of GIS to the school system around the world already began at the beginning of the twenty-first century. Analyzing the current use of geographic information systems in schools around the world, it should be noted that the US historical leadership in the development of GIS education was in the undoubted leadership of the use of GIS in secondary schools in the country. Undoubtedly, ESRI's policy has been the main factor in this, which has made GIS education of its highest priority in schools, colleges and universities [7].

In this article, we would like to draw the attention of the educational community to the potential of ArcGIS Online Clouds offered by ESRI.

ArcGIS Online is a complete geographic information system hosted on a cloud-based server with a broad functionality. With ArcGIS Online, you can create web-maps, use ready-made resources, publish mapping services, spatial analysis, distribute data, and access cards from any device. At the same time, ArcGIS Online can be used as a platform for building your own geographically-bound applications. Through the built-in map viewer, galleries of base maps and space images are provided free of charge, and their range, detail and relevance are constantly expanding [1; 2].

If you add to this the benefits of cloud technologies, such as access to personal information from any computer, reliable cloud storage, a significant reduction in the cost of purchasing licensed software, as well as the technology of distributed data processing in which the computer resources and capacity are provided to the user as an Internet service, then it becomes clear that ArcGIS Online is a great tool for working with geospatial data at different stages of future geography teachers' training.

It should be borne in mind that the criterion for motivating the use of GIS-technologies is their influence on the students mastering the material in different conditions. In this regard, in each case, the purpose of the GIS application and its functional definition must be clearly specified. It is expedient, the use of this technology can only be considered reasonably motivating if pedagogical efficiency cannot be achieved using other, more accessible means of study.

Let us determine the basic principles of using ArcGIS Online in an educational process:

1. *Interdisciplinary integration.* Today, maps, tables, graphs and charts – all these kinds of representations of information generated by GIS are widely used in disciplines related to geography: chemistry, history, computer science, biology, mathematics, etc. It allows students to integrate knowledge from these disciplines. Integration occurs through work with electronic maps and databases, the topics of which are related to the discipline, and the tool is GIS.

For example, the use of students' knowledge in mathematical statistics helps to identify causal relationships between different natural components of the geographical environment. Knowledge in chemistry is essential to highlight issues of the migration of chemical elements during environmental research. The use of knowledge in biology reveals the interrelations between elements of natural landscapes. ArcGIS Online allows one to combine these diverse data into a single geospatial model. The use of GIS is especially encouraged by the development of computer literacy for future geography teachers: file management, database work in the cloud, satellite positioning systems, remote sensing data, etc.

2. *The principle of consistency* ensures the gradual and promising learning of cloud GIS technologies; in other words, the assimilation of knowledge takes place from simple to complex. The educator should not speed up the process of familiarizing and mastering ArcGIS Online by students. Only competently organized, step-by-step and metered training is the key to a successful work with this service.

So, at the initial stage, students should familiarize themselves with the cloud-based GIS interface, log into the system and create a user's personal page. At the next stage of training, future educators will get acquainted with online map creation, work with layers and attributes, which will eventually lead to the stage of registration and editing of thematic maps and organization of the general access to the elements.

As a result, students must independently create cartographic web-based applications. The development of cartographic web-applications for learning purposes is based on the use of ready-made templates in the cloud, which are a convenient way of publishing web-maps and combine location with web-based applications, as well as provide multimedia and interactive features.

3. *Individualization of training.* Work in the cloud refers to personality-oriented technologies that help the teacher to individualize the process of learning, to organize an individual educational process for each student. This significantly transforms the activities of the subjects of the educational process – the teacher and students. They have to engage in fundamentally new activities in connection with the change of educational activities and the specific restructuring of its content.

Individualization of training with GIS involves the differentiation of the teaching material, the system of tasks of varying complexity and volume. It is advisable to highlight the main and varied study materials. It is expedient to actively use GIS-projects, with their division into separate small tasks, stages. In addition, each subsequent task becomes feasible for the student, providing the previous one has been accomplished [7].

4. *The principle of communicability.* Observance of this principle involves such class orientation, when the purpose of learning (mastering GIS-technologies) and the means of achieving the goal (spatial and temporal analysis in the study of natural, socio-economic, man-made processes, objects and phenomena) are closely interrelated. First of all, this involves activating the students' creative activity: the widespread use of collective forms of work, problem situations, creative tasks, etc. Cloud technology is still more likely to be conducive in the organization of joint activities and interaction of subjects of the educational process, the ability of future educators to take into account different opinions and strive for the coordination of different positions.

Each student, at the time of registration in the service ArcGIS Online, gets their own design site, where they store their own attributes and data, maps and other useful information.

5. *Principle of distance learning.* By implementing cloud-based GIS technologies in training, students are not required to have a physical presence at the place of their education. This technology allows students to use educational materials of any kind, as well as perform work with teachers or a group. ArcGIS Online allows teachers and students to share their research results through the use of supported GIS data and services, thus creating an information space that other users can access.

Obtained thematic maps can be published on the Internet as content pages or as web-based applications using built-in mapping templates for electronic maps. Among these templates, attention is drawn to the ones with synchronized windows, which allow you to view various subject-specific data in terms of a single territory.

6. *The local area study principle* is realized in the practical orientation and research competence of GIS-technologies. On the one hand, it determines a comprehensive study of this or that local area, and on the other hand, the use of ethnographic material in teaching. Both of these tasks are interrelated: the first solution opens the way to the second.

The regional orientation of cloud GIS can be engaged in a wide range of geographic research areas: mapping and plans of its terrain, analyzing the meteorological situation, studying the spatial distribution of biological species, socio-ecological research. The material required for conducting research using GIS can be obtained by applying literary, cartographic and statistical methods, as well as by method of direct observations during the local area studies, travels and excursions.

An interactive map is the main form of research for future geography teachers through the help of cloud-based ArcGIS Online. The main functions of this online service provide a simple and accessible form of geospatial information visualization in the form of interactive maps (Figure 1), as well as their publication on the Internet.

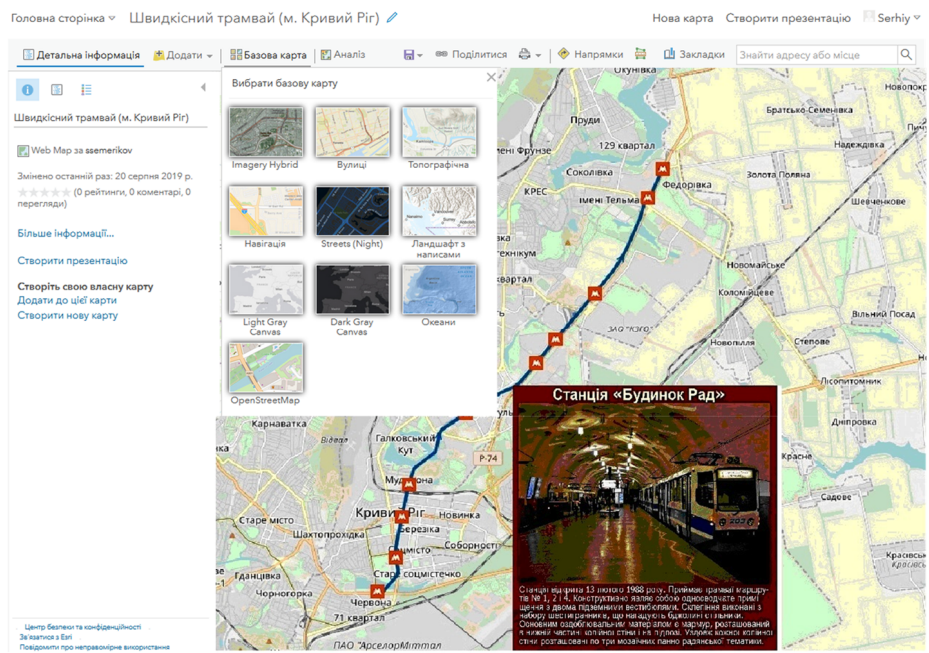

**Fig. 1.** The view of the interactive map "High-speed tram at Kryviy Rih", designed by means of using ArcGIS Online cloud

The use of innovative teaching aids requires the teacher to master new forms and methods of activity. The integration of such high-tech information media as cloud GIS into the educational process should be based on a clear pedagogical strategy:

1. Cloud GIS, as interactive learning tools, allows teachers to manage the flow of information, focusing on the most interesting or more complex moments of the material being studied, allowing them to model geographic phenomena and processes, demonstrating them dynamically. Consequently, they facilitate the understanding of the essence of these phenomena and processes.
2. In the learning process, the teacher should not act as a software expert. ArcGIS Online is a rather complex cloud service, and in this connection it is impossible to know all its features. Therefore, the teacher should constantly be engaged in self-education.
3. The teacher should not be afraid to change pedagogical approaches to learning using cloud GIS. He must be in the creative search for new ideas, be prepared to make adjustments to the teaching methodology taking into account both the dynamics of the level of students' general knowledge as well as their level of preparation in geoinformation.
4. Re-perform is the key to the success of GIS education. If the required function is not provided for some reason, it is actually useful, because it forces the student to go through this path over and over again, fixing the algorithm of the task in their brain.
5. An important task for a teacher when using ArcGIS Online is to create a competitive learning environment among future educators as it significantly enhances their motivation. It is important that, in selecting forms and methods for organizing a competitive learning environment, the teacher takes into account the personal characteristics of the students.
6. When using cloud-based GIS in the training of geography teachers, the teacher should share his or her experience among the colleagues, both during direct consultation and in specialized seminars, conferences, webinars, online forums, electronic and print magazines, etc. On the one hand, it will allow them to distribute the results of their own work, and on the other hand – take into account the progress of other teachers, avoiding unnecessary mistakes in their own activities.

Like traditional maps, ArcGIS Online is a three-way learning tool: a means of visibility, an object of study, and a source of knowledge. Accordingly, the following options for using GIS in the educational process are available:

1. *Work in the demonstration mode.* Cloud GIS have a unique ability to create a visual spatial image of various objects, processes and phenomena, and the teacher should be able to use it. ArcGIS Online has a huge number of basic maps that the teacher can use in his or her work. The gallery contains a variety of topographical, demographic, socio-economic maps and satellite images of the world. In addition, these maps can be used as a basis for creating your own maps.

The layer-by-layer form of information organization, the ability to complement cartographic information by various schemes and charts, demonstration of dynamic processes, 3-D models and much more, allow the teacher to form a new way in the students' geographic concepts during the study of new material.

2. *Work in a computer class.* The conduct of practical classes or research should be regarded as the most effective work with cloud computing ArcGIS Online in the computer class. The following activities are available in terms of content and proposed methods of organization:

- frontal work, carried out by the whole group, when all students perform the same task;
- group work, for which students are divided into small groups (3-5 people);
- pair work, when two are studying a question;
- individual work, carried out by the student himself or herself.

3. *Work in an individual mode.* The individual work of future geography teachers with cloud GIS is possible mainly in the form of distance learning organization of the educational process, which is aimed at the development of the students' personality, their autonomy, creativity. The most effective form of training in this case should be recognized as a project activity.

The project method is to create conditions for independent mastering of cloud-based GIS-technologies by geography students during the realization of specific projects. They participate in this process starting from the project idea itself up to its practical implementation. As a result, students, with the help of a teacher, learn to search, summarize and analyze information independently, enter it into a cloud database, build their own map sets.

The themes of projects should be determined by the sphere of interests of future teachers and depend on the level of their individual training. So, in the first stages, it may be some task that demonstrates the relationship between cartographic and attribute information. These are various information GIS projects, where information about different historical, social and natural objects of the studied territory is displayed. In the future, the teacher should focus on projects that use the analytical functions of GIS (over-lined operations, buffering, zoning, etc.).

The main advantages of cloud GIS as an educational tool are:

- round-the-clock access to personal information from any computer connected to the Internet;
- no need to install a GIS package on your computer;
- access to spatial resources (various maps from physical and geographic to socio-economic and political, satellite images, historical maps, interactive maps, etc.) prepared and published by other users;
- personal data storage in the cloud: map packs, layers, tabular (attribute) data, etc.;
- the possibility of creating interactive maps, as well as creating web-based applications based on pre-made templates without the use of programming tools;
- ease of cooperation integration between the users within an educational organization.

At the same time, it is necessary to describe the problems associated with the widespread use of ArcGIS Online in geography teacher training:

- monetization of the resource: the full-fledged use of the resource is possible only on a paid basis, and not all educational institutions of higher education can afford it;
- a small number of class hours in the curriculum of future geography educators, dedicated to the study of GIS-technologies and their full-fledged mastery;
- absence of the possibility of joint editing of web-maps in an on-line mode;
- absence of Ukrainian-language content for the resource.

## 3    Conclusions

1. To summarizing, it should be noted that cloud GIS is a new stage in the development of geoinformational education. Thanks to them, work with GIS is more efficient and affordable. Cloud ArcGIS Online has proven to be a fairly simple, well-developed and user-friendly educational resource.
2. The prospects for further research are seen in the formation of training and methodological base for integrating the elements of cloud GIS into various training courses for future geography teachers.